\documentclass{pasj01}
\usepackage{color}

\begin{document}
\SetRunningHead{Shogo B. KOBAYASHI}{Shogo B. KOBAYASHI}

\title{Suzaku Observations of Spectral Variations of the Ultra Luminous X-ray Source Holmberg IX X-1}

\author{Shogo B. \textsc{KOBAYASHI}\altaffilmark{1}}
\altaffiltext{1}{Department of Physics, The University of Tokyo, 7-3-1 Hongo, Bunkyo-ku, Tokyo 113-0033}
\email{kobayashi@juno.phys.s.u-tokyo.ac.jp}

\author{Kazuhiro \textsc{NAKAZAWA}\altaffilmark{1}}

\author{Kazuo {\textsc MAKISHIMA}\altaffilmark{2}}
\altaffiltext{2}{MAXI team, RIKEN, 2-1 Hirosawa, Wakou-shi, Saitama-ken 351-0198, Japan}

%


\maketitle

\begin{abstract}

Observations of the Ultra Luminous X-ray source (ULX) Holmberg IX X-1 were carried out with Suzaku twice, once on 2012 April 13 and then on 2012 October 24, with exposures of 180 ks and 217 ks, respectively. The source showed a hard power-law shape spectrum with a mild cutoff at $\sim 8$ keV, which is typical of ULXs when they are relatively dim. On both occasions, the $0.6\--11$ keV spectrum was explained successfully in terms a cool ($\sim 0.2$ keV) multi-color disk blackbody emission model and a thermal Comptonization emission produced by an electron cloud with a relatively low temperature and high optical depth, assuming that a large fraction of the disk-blackbody photons are Comptonized whereas the rest is observed directly. The $0.5\--10$ keV luminosity was $1.2\times10^{40}$ erg s$^{-1}$ in April, and $\sim 14\%$ higher in October. This brightening was accompanied by spectral softening in $\ge 2$ keV, with little changes in the $\le 2$ keV spectral shape. This behavior can be understood if the accretion disk remains unchanged, while the electron cloud covers a variable fraction of the disk. The absorbing column density was consistent with the galactic line-of sight value, and did not vary by more than $1.6\times 10^{21}$ cm$^{-2}$. Together with the featureless spectra, these properties may not be reconciled easily with the super-critical accretion scenario of this source.
\end{abstract}

\section{Introduction}
Ultra Luminous X-ray sources (ULXs; Makishima et al~2000) are unusually luminous X-ray point sources that are often found at places of enhanced star formation, e.g., arms of spiral galaxies. Their luminosity, ${\it{L}}_{\rm{x}}\sim10^{39.5-41.5}$ erg s$^{-1}$, significantly exceeds the Eddington limit of a stellar mass ($\sim10\ {\it{M}}_{\odot}$) black hole (BH), ${\it{L}}_{\rm{Edd}}=1.5\times 10^{39}$ ($M/10 \ M_{\rm{\odot}}$) erg s$^{-1}$, where $M$ is the BH mass and $M_{\rm{\odot}}$ is the solar mass. Since their first discovery with the Einstein Observatory (Fabbiano et al.~1989), ULXs had remained mysterious objects for more than two decades. When the wide band spectroscopy of ULXs became available with ASCA, their spectra turned out to be explicable with the multi-color disk (MCD; Mitsuda et al.~1984, Makishima et al.~1986) model, which represents multi-temperature blackbody radiation from a standard accretion disk around accreting objects (Makishima et al.~2000). This implies their close resemblance to ordinary BH binaries (BHBs). In order to reconcile their high luminosities with the relatively high inner-disk temperature ($T_{\rm{in}}\sim 1$ keV) obtained with the MCD-model fitting, Makishima et al.~(2000) proposed that ULXs are intermediate mass ($100-1000 \ M_{\rm{\odot}}$) BHs with significant angular momentum, radiating at or somewhat below than their Eddington luminosities. 

The existence of such relatively heavy ($\ge 20 \ M_{\rm{\odot}}$) BHs has been clearly demonstrated by the epoch-making detection of the gravitational-wave event, GW 150914 (Abbott et al.~2016), in which two BHB with $\sim 30 M_{\rm{\odot}}$ merged into one with $\sim 60 M_{\rm{\odot}}$. Although the formation scenario of these $\sim 30 M_{\rm{\odot}}$ BHs in binaries is yet to be identified (cf.  Ebisuzaki et al.~2001; Kinugawa et al.~2014), the merger remnant could eventually become a ULX, even without a mass-donating companion, through direct accretion from thick interstellar matter (Mii \& Totani 2005; section 2). This gives a support to the interpretation of ULXs in terms of intermediate-mass BHs.

In paralllel with the above approach, the puzzling properties of ULXs motivated alternative interpretations to be proposed and extensively developed. For example, King et al.~(2001) argued that ULXs are stellar-mass BHs with their emission highly collimated towards us. Similarly, some authors (e.g. Watarai et al.~2000; Mineshige \& Ohsuga~2007) pointed out that the Eddington limit can be surpassed in an accretion disk system, which is predominantly axi-symmetric, and adopted an accretion disk model called Slim-Disk model to explain the MCD-like spectra. This model describes emission from optically-thick super-critical accretion flows with strong advection, and predicts MCD-like spectra with considerably higher temperatures. They hence concluded that the high $T_{\rm{in}}$ spectra are due to super critical accretion flows, and ULXs harbor stellar-mass BHs instead of intermediate ones. Actually, this viewpoint has become very popular both observationally and theoretically (e.g., Middleton et al.~2011, Ohsuga \& Mineshige~2011, Sutton et al.~2013, Middleton et al.~2015). In any event, it still remains controversial whether ULXs are considerably more massive than the stellar-mass BHs, or are in the state of significant super-Eddington emission (either apparently or intrinsically).

Thanks to extensive studies with many X-ray observatories, ULXs have been found to show two characteristic spectral states (e.g. Gladstone et al.~2009, Pintore et al.~2014, Luangtip et al.~2016). One is the MCD-like state mentioned above, while the other is Power-Law (PL) like state. Although the PL-state spectra were at first represented by a single PL, two features were noticed as the observation made a progress. One is a mild cutoff above $8\-- 10$ keV (Kubota, Done \& Makishima~2002), suggestive of thermal Comptonization (THC) process. The other is \textquotedblleft soft excess\textquotedblright\ often seen below 1.5 keV of these spectra
(Gladstone et al.~2009), which suggests the presence of an optically-thick emission that is cooler ($\sim0.3$ keV) than those describing the MCD-state spectra. In fact, the X-ray spectra of these ULXs were well reproduced by assuming a corona that is partially covering an optically-thick source. Specifically, it is assumed that some fraction of the optically-thick emission is directly visible as the soft excess, and the rest is up-scattered via the THC to form the PL continuum. The optically-thick emission could be coming from an accretion disk (MCD+THC modeling; Dewangan et al.~2006, Goad et al.~2006, Gladstone et al.~2009, Miyawaki et al.~2009, Pintore et al.~2014), or alternatively from other sources such as photosphere of optically thick outflows (Middleton et al.~2011).

While the above MCD+THC model is similar to a modeling which is often used in BHBs (Kubota et al.~2001, McConnell et al.~2002), the ULX spectra exhibit significantly lower coronal temperature, $2\-- 3$ keV, compared to those of BHBs in the Low/Hard states ($\sim 100$ keV; e.g. Makishima et al.~2008, Yamada et al.~2013) or Very High states ($\sim$ few tens keV; e.g. Kubota \& Done~2004). Some works on computational simulations suggest that these low temperature coronae can be identified with optically thick cool outflows launched from super-critical accretion flows onto stellar-mass BHs (e.g. Ohsuga et al.~2003, Kawashima et al.~2012), which should bear rich features of X-ray reprocessing, including photoelectric absorption, fluorescence lines, and resonant absorption lines (e.g. Hagino et al.~2016). However, the actually observed ULX spectra are extremely featureless, posing another puzzle. 

To better understand ULXs, we examine whether the PL-like spectra are explicable as an extension of ordinary sub-Eddington modeling developed through the past BHB studies. Specifically, we apply the MCD+THC modeling to some ULX spectra and examine the results for consistency with those on BHBs. Since our study requires as high statistics as possible, we selected in the present paper the representative ULX, Holmberg (Hol) IX X-1, which is one of the brightest of these X-ray sources.

\section{Observation}
Hol IX X-1 is a ULX in the irregular dwarf galaxy Hol IX, which is associated with the nearby spiral galaxy M81 (NGC 3031) at a distance of 3.4 Mpc (Georgiev et al.~1991). It is the brightest X-ray source ($L_{\rm{X}}\sim1\--3\times 10^{40}$ erg s$^{-1}$ at $0.5\--10$ keV) within $20'$ of the M81 nucleus except the nucleus itself, and has been studied extensively. Its spectra obtained by ASCA and XMM-Newton, in 1998 and 2001 respectively, were explained with emission from an accretion disk with a high inner temperature as $\sim 1.3$ keV (Tsunoda et al. 2006, La Parola et al. 2001); these data provide typical examples of the MCD-like spectra of ULXs. After these observations, the source has in contrast been residing in the PL state, wherein the spectrum can be explained with the MCD+THC modeling (e.g. Dewangan et al.~2006, Vierdayanti et al.~2010) as described in section 1. 

In an attempt to detect any evidence for outflows, Walton et al.~(2013) conducted the historically deepest observation of Hol IX X-1 in 2012 with Suzaku, utilizing its higher energy resolution and lower background around $6.4$ keV than any other X-ray observatories operating at that time. In spite of this effort, they failed to detect significant local spectral features, and obtained a stringent upper limit of 30 eV for any narrow absorption or emission line around $6.4$ keV (Walton et al.~2013). In the present paper, we re-analyzed the same Suzaku data to study the variability of spectral continuum of Hol IX X-1.

As already described by Walton et al.~(2013), the observation was carried out on 2012 April 13-17 for a net exposure of 183 ks, and again on 2012 October 24-28 for 217 ks. The $\sim6$ months separation between the two observations allows us to study possible long-term variations of the source. In both these observations, the X-ray Imaging Spectrometer (XIS) and the Hard X-ray Detector (HXD) were operated normally. Front-Illuminated (FI) CCD cameras (XIS0 and XIS3) of the XIS, and the Back-Illuminated (BI) one (XIS1), were both operated in their standard modes, with charge injection, but without window option. The source was placed at the \textquotedblleft XIS nominal\textquotedblright\ position in the XIS field of view. Since the HXD data were heavily contaminated by X-rays from the M81 nucleus (Dewangan et al.~2013), we utilize only the XIS data in the present paper. 

\begin{figure}
  \begin{center}
    \FigureFile(80mm,80mm){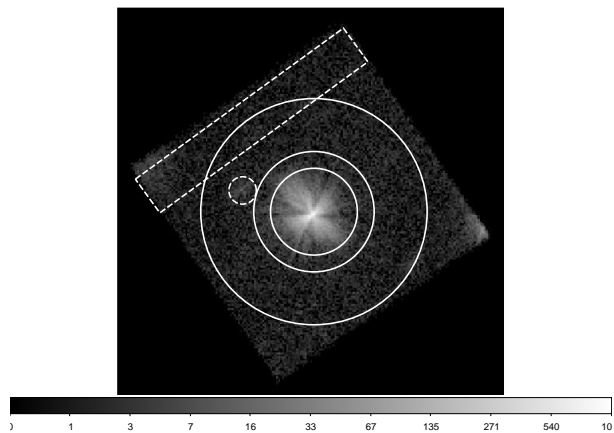}
  \end{center}
  \caption{Background-inclusive $0.5\--10$ keV XIS3 image of Hol IX X-1. The innermost circular region was used to derive the light curves and the on-source spectra. The annular region between the largest and middle circles was used to obtain background spectra. Two regions in dashed line were excluded (see text).}
  \label{fig:xisimg}
\end{figure}

\section{Data Analysis and Results}
\subsection{XIS Image}
Figure \ref{fig:xisimg} gives the background-inclusive $0.5-10.0$ keV image of XIS3. The source region was thus defined as a $2'.8$-radius circle centered on the source, and the background region as an annulus of the inner and outer radii of $3'$ and $8'$, respectively. We excluded a rectangular region in figure \ref{fig:xisimg} where some dead columns exist in XIS0, and a small circle region where a faint X-ray point source is present. At the right bottom of the field of view, we notice some stray light from the M81 nucleus.

\subsection{XIS Light Curves}

\begin{figure}[t]
 \begin{center}
   \FigureFile(80mm,80mm){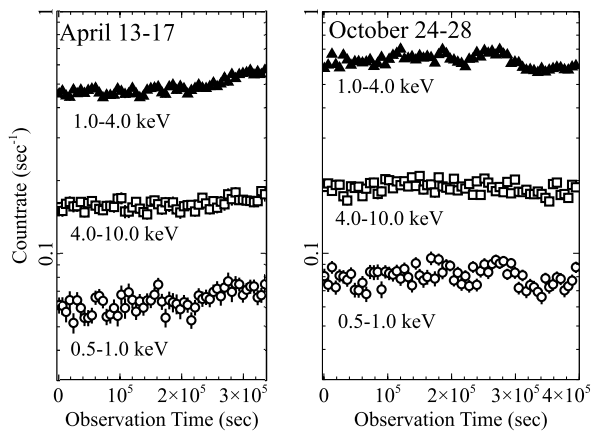}
  \end{center}
    \caption{Background-subtracted XIS0+XIS3 light curves from the April (left) and October (right) observations, in energies of $0.5\--1.0$ keV (open circles), $1.0\--4.0$ keV (filled triangles), and $4.0\--10.0$ keV (open squares). Each bin is set to be the same as the satellite orbital period of 5.76 ks.}
    \label{fig:xislc}
\end{figure}
Figure \ref{fig:xislc} shows background-subtracted XIS light curves from the two observations in three energy bands, derived using the signal and background regions defined in figure \ref{fig:xisimg}. The gross pointing thus lasted for $\sim 340$ ks in April and $\sim 400$ ks in October. As usual with Suzaku observations, the net exposure is about $40\--50 \%$ of these values. While the hard-band ($4.0\--10.0$ keV) light curve was nearly constant within $\sim 10\%$ throughout each observation, the others showed somewhat more enhanced variability. In addition, the average count rate increased from April to October by $27 \%$, $28 \%$, and $17 \%$, in the soft ($0.5\--1.0$ keV), middle ($1.0\--4.0$ keV), and the hard ($4.0\--10.0$ keV) band, respectively. The actual flux increase (section 3.3), however, is somewhat smaller then these, because the source position was slightly different between the two observations.

\subsection{XIS Spectra of Hol IX X-1}
\begin{figure}[h]
	\begin{center}
		\FigureFile(80mm,80mm){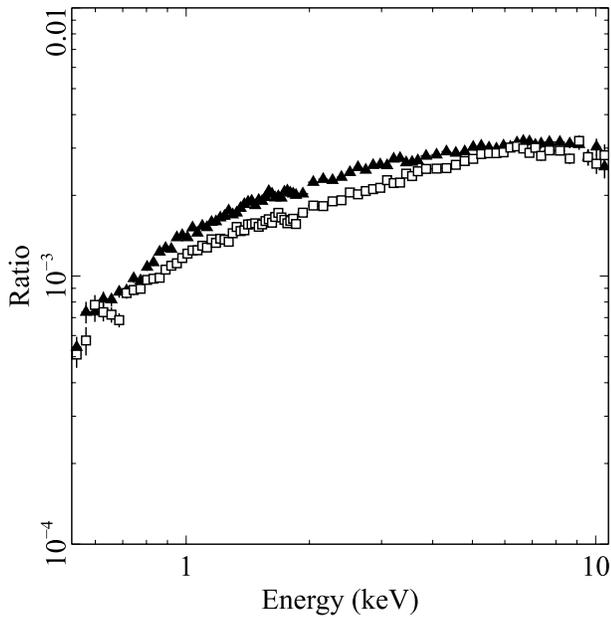}
	\end{center}
		\caption{XIS 0+3 Spectra obtained in the April (open squares) and the October (filled triangles) observations. The spectra are shown in ratios over a common PL model with a photon index of 2.}
		\label{fig:powerratio}
\end{figure}

\begin{figure}[h]
 \begin{center}
   \FigureFile(80mm,80mm){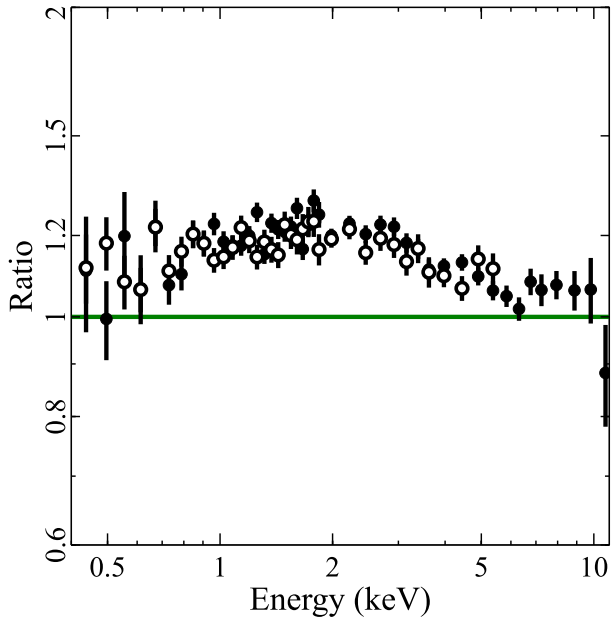}
  \end{center}
    \caption{The XIS spectra in October, divided by those in April. Filled and open circles represent the FI XIS and the BI XIS data, respectively. Effects of the slight difference in the pointing position are neglected.}
    \label{fig:monthly}
\end{figure}
The same regions in figure \ref{fig:xisimg}, as used to produce the light curves, were again employed to extract on-source and background spectra. In order to maximize the statistics, we summed the FI XIS (XIS0 and XIS3) spectra together, because they have almost the same instrumental response. The background-subtracted spectra obtained from the individual observations are presented in figure \ref{fig:powerratio}. In order to grasp approximate shapes of the spectra, we took their ratios to a PL model with photon index of 2. For simplicity, only the FI XIS spectra are shown, but those with BI XIS are consistent. Both spectra thus exhibit hard PL shapes with a turn over at $\sim 8$ keV, and a low energy break at $\le 1$ keV due to absorption by intervening matter. As expected from the light curve, the $1\-- 4$ keV flux increased in the 6 months, and made the spectrum somewhat softer in the October observation.

In order to study the characteristics of this spectral variability in more detail, but without using any physical model, we took ratios between the spectra on the two occasions, and present the result in figure \ref{fig:monthly}. Here, we also show the result from the BI XIS, which is consistent with that from the FI XIS. The ratio is relatively flat in $\le 2$ keV, indicating that no significant change took place in the absorbing column density, and the spectral shape in $\le 2$ keV remained unchanged as the luminosity varied. On the other hand, the data points in $\ge 2$ keV form a straight PL form up to $\sim7$ keV, where the spectra already start bending (figure \ref{fig:powerratio}). This suggests that the cutoff energy remained relatively unchanged between the two spectra, and mainly the slope changed. 

\subsection{Model Fitting}
\subsubsection{Individual Fitting}
\begin{figure*}[pt]
 \begin{center}
   \FigureFile(160mm,320mm){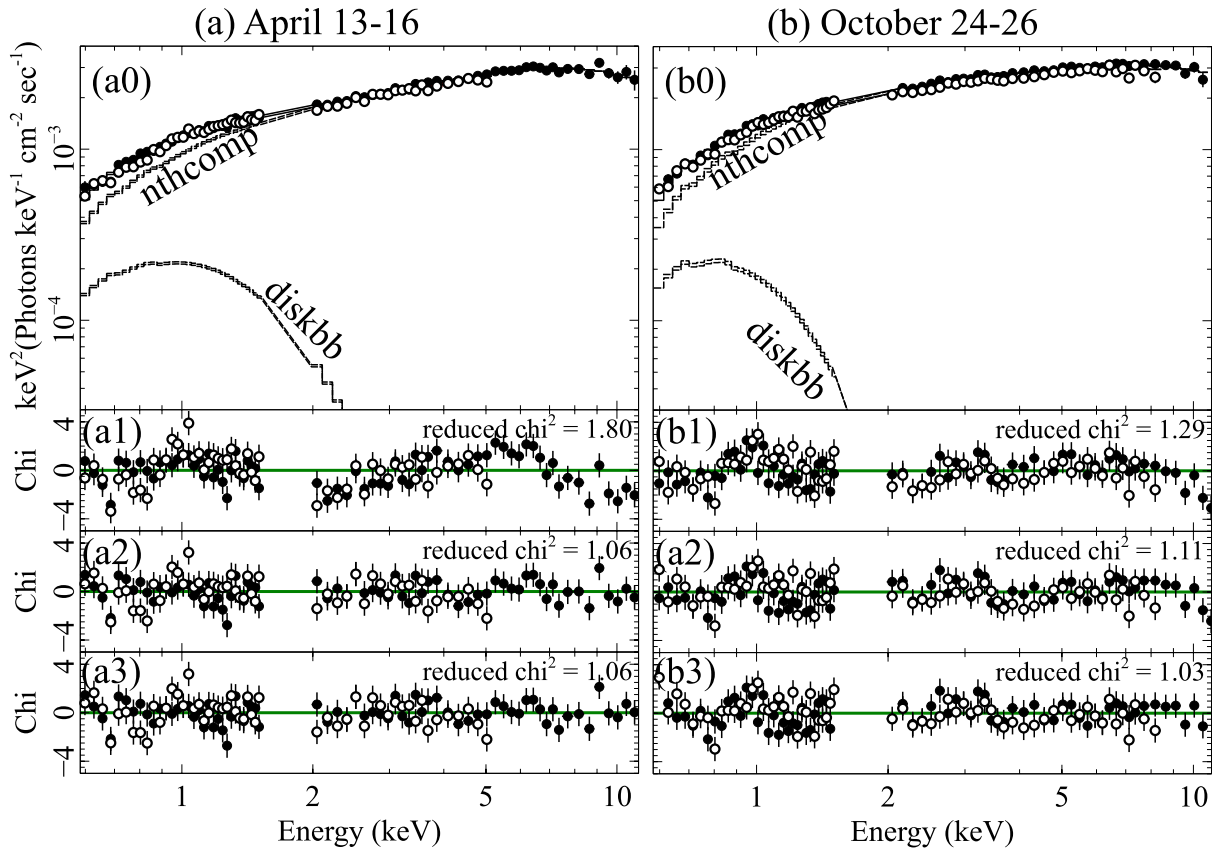}
   \end{center}
    \caption{Suzaku XIS spectra for the two observations (panel a0 and b0), and their residuals from the utilized models (panels a1-a3 and b1-b3). The April and October observations are distinguished by \textquotedblleft a\textquotedblright\ and \textquotedblleft b\textquotedblright\, respectively. Panels (a1) and (b1) are residuals from Model 1, (a2) and (b2) from Model 2, while (a3) and (b3) from Model 3. Data points from FI XIS and from BI XIS are plotted in filled and open circles, respectively. The model components illustrated in (a0) and (b0) refer to those from Model 3.}
    \label{fig:apavesp}
\end{figure*}

As presented in figure \ref{fig:apavesp}, the derived XIS spectra were quantified via model fitting. The April and October data were analyzed separately, whereas the XIS FI and BI data were fitted jointly. Since BI XIS suffers higher background, we used a more limited energy band for its spectra ($0.6\--5.0$ keV in April, and $0.6\--8.0$ keV in October) than for the XIS0+3 data ($0.6\--11.0$ keV). In addition, we excluded the $1.5\-- 2.0$ keV band from our analysis, and multiplied a constant factor to the XIS1 spectrum. These are intended to deal with uncertainties in the instrumental response around the Si edge, and in the cross calibration among two type of CCDs, respectively. Our analysis was carried out by using the XSPEC ver 12.7.1.

 Instead of immediately jumping onto the MCD+THC modeling described in section 1, we tried several spectral models with progressive complexity. Based on the approximately PL-like spectral shape (figure \ref{fig:powerratio}), we first employed a cut-off PL ($\tt{cutoff pl}$) model, multiplied by photoelectric absorption model $\tt{phabs}$; this is called Model 1. We allowed the column density $N_{\rm {H}}$ of the absorber to float. The fitting results are shown in table \ref{tb:avefitp} for this and subsequent models, where the error refers to the $90 \%$ confidence limits of each parameter. 
 
 As shown in panels (a1) and (b1) of figure \ref{fig:apavesp}, the spectra were approximately explained with this model, by photon indices of $\Gamma =1.6\-- 1.7$ and cutoff energies of $E_{\rm{cut}}= 23 \--43$ keV. However, the fit leaves some positive residuals in $0.8\--1.5$ keV, which suggest that some fraction of emission from the accretion disk is reaching us directly. Therefore we added an MCD component ($\tt{diskbb}$) to the model, thus constructing Model 2. This $\tt{diskbb}$ component has two free parameters; the inner-disk temperature $T_{\rm{in}}$ and the apparent inner-disk radius $R_{\rm{raw}}$, which are mainly determined by the peak energy of the soft excess component, and its flux, respectively. As a result, the fit became acceptable in both data sets (table \ref{tb:avefitp}). The $0.5\--10$ keV luminosity increased by $\sim 14 \%$ from April to October as given in table \ref{tb:avefitp}. Since the soft excess is weak in the spectra (figure \ref{fig:apavesp}), $\tt{diskbb}$ is subject to rather large errors.
 
 Since the cutoff-PL-shape continuum often arises via Comptonization, we replaced $\tt{cutoff pl}$ with a THC code called $\tt{nthcomp}$ (Zdziarski, Johnson \& Magdziarz~1996, Zycki, Done \& Smith~1999). The seed-photon source of $\tt{nthcomp}$ was chosen as another $\tt{diskbb}$, of which the temperature is set identical to that of the directly visible $\tt{diskbb}$. In other words, we assume that the source has a single disk that is partially covered by the corona. This form of model, to be called Model 3 hereafter, was employed by Gladstone et al.~(2009) in their study of ULXs, and is frequently used in BHB studies to represent the THC process of the seed photons from a disk (e.g. Makishima et al.~2008). The model determines the coronal temperature $T_{\rm{e}}$ and the photon index $\Gamma$ from the cutoff and the slope of the spectrum, respectively. The model has successfully reproduced the two spectra, with a similar fit goodness to that with Model 2. The derived parameters are again listed in table \ref{tb:avefitp}, where we converted, after Sunyaev and Titarchuk ~(1980), the obtained $T_{\rm{e}}$ and $\Gamma$ into the coronal optical thickness $\tau$ as, 
 \begin{equation}
 \tau = \left\{2.25+\frac{3}{(kT_{\rm{e}}/511\ \rm{keV})[(\Gamma + 0.5 )^{2}-2.25]}\right\}^{-\frac{1}{2}}-\frac{3}{2} \ .
 \label{eq:tau}
 \end{equation}
Since the spectrum in October exhibits a softer (larger) $\Gamma$ than that in April, equation (\ref{eq:tau}) gives a slightly smaller value of $\tau$ (table \ref{tb:avefitp}) that in the former spectrum.

\begin{table*}[htp]
\caption{Results of the model fits to the April and October spectra.}
\begin{tabular}{ccccccccccc}
  \hline
  Model 	& $N_{\rm{H}}$$^{\ast}$	&	$\Gamma$$^{\dagger}$ 	& $T_{\rm{e}}$ or $E_{\rm{cut}}$ & 	$\tau$$^{\ddagger}$ 	&$T_{\rm{in}}$ 			&$R_{\rm{raw}}$$^{\S}$ 	& $R_{\rm{in}}$$^{||}$	& $L_{\rm{disk}}$$^{\#}$	&$L_{\rm{X}}$$^{\ast \ast}$	&	red. $\chi^{2}$ ($\nu$) 	\\
  \hline
  April	& 					&						& 							& 						& 						& 					& 					&					&						&						\\
  1  	&	$0.91\pm0.09$	&	$1.57\pm0.04$		&	$43^{+43}_{-15}$ 			& 	$...$					&	$...$					& 	$...$ 			& 	$...$				& 	$...$				&			$1.25$		&	$1.80 \ (119)$		\\
  2		&$0.94^{+0.02}_{-0.03}$	&	$0.98^{+0.17}_{-0.22}$	&	$7.3^{+2.3}_{-1.6}$			&	$...$					&	$0.37\pm 0.06$		&  $750\pm290$	& 	$...$				&	...				&			$1.24$		&	$1.06 \ (117)$		\\
  3		&	$1.0\pm0.3$		&	$1.65^{+0.02}_{-0.03}$	&	$2.71^{+0.20}_{-0.17}$		&	$\sim 14$			&	$0.30^{+0.06}_{-0.05}$	&  $850^{+560}_{-670}$ &	$2480$			& 	$6.4$			&			$1.24$		&	$1.06 \ (117)$		\\
  \hline
  October	&					&						&							&						&						& 					& 					& 					&						&						\\	
  1		&	$1.12\pm0.08$	&	$1.61\pm 0.04$		&	$23^{+7}_{-5}$				&	 $...$				&	$...$					& 	$...$ 			& 	$...$				&		$...$		&			$1.42$		&	$1.29 \ (131)$		\\
  2		&	$1.3\pm0.3$		&	$1.43^{+0.10}_{-0.15}$	&	$12^{+4}_{-3}$				&	 $...$				&	$0.28^{+0.10}_{-0.06}$	& $1030^{+960}_{-490}$			& 	$...$				& 		$...$		&			$1.41$		&	$1.11 \ (129)$		\\
  3		&	$1.8^{+0.5}_{-0.6}$	&	$1.79\pm 0.02$		&	$3.2^{+0.4}_{-0.3}$			&	 $\sim 12$			&	$0.18^{+0.04}_{-0.02}$	& $3710^{+2680}_{-2540}$ 		& 	$9700$		& 	$12.7$		&			$1.41$		&	$1.03 \ (121)$		\\
  \hline
  \end{tabular}
  \begin{flushleft}
  	$\ast$: Column density of equivalent Hydrogen in unit of $10^{21}$ cm$^{-2}$.\\
	$\dagger$: Photon index of the power-law component.\\
	$\ddagger$: Optical depth of the electron cloud.\\
	$\S$: Apparent inner radius of the seen-through (un-scattered) accretion disk component in units of km.\\
	$||$: Innermost radius of the overall accretion disk in unit of km; equation (\ref{eq:rin}).\\
	$\#$: Bolometric luminosity of the overall accretion disk in unit of $10^{39}$ erg s$^{-1}$.\\
	$\ast\ast$: $0.5\--10$ keV band absorbed luminosity in unit of $10^{40}$ erg s$^{-1}$. \\
  \end{flushleft}
  \label{tb:avefitp}
\end{table*}

\subsubsection{Simultaneous Fitting}
Although Model 3 has provided successful fits to the spectra on both observations, the derived parameters except $\Gamma$ do not differ significantly between the two data sets. To better characterize the spectral variability revealed by figure \ref{fig:monthly}, we fitted the April and October spectra simultaneously with some parameters tied together, and obtained the results summarized in table \ref{tb:simfitp}. That is, we began with the condition that all the parameters are allow to differ between the two spectra, and gradually tightened their constraints. When all the parameters are allowed to be free and independent (ID 1), we naturally obtained the same results, within errors, as those in Model 3 of table \ref{tb:avefitp}. 

According to figure \ref{fig:monthly}, the spectral ratio is flat below $\sim2$ keV, and shows a straight form up to $\ge 8$ keV. These suggest that the absorption and the cut-off energy both remained unchanged across the 6 months which separate the two observations. Therefore, we next constrained $N_{\rm{H}}$ and $T_{\rm{e}}$ both to be the same between the two data sets (ID 2). Then, $\chi ^{2}$ increased by 7 for an increase of 2 degrees of freedom, but the fit still remained acceptable at a similar statistical level to that of ID 1. Since we fixed $N_{\rm{H}}$ to a single common value, the disk component now accounts for the variability below 2 keV. 

To further confirm the suggested disk variation, we tied all the parameters of the disk component, $T_{\rm{in}}$ and $R_{\rm{raw}}$, to be the same between the two spectra, and let only $\Gamma$ differ between them (ID 3). This made the fit unacceptable, increasing $\chi ^{2}$ by 23 for an increase of 2 degrees of freedom from what we obtained in ID 2. Thus, in addition to $\Gamma$, either $T_{\rm{in}}$ or $R_{\rm{raw}}$ or both must have changed. Then, we need to know which of them accounts for the variability.

\begin{figure}[htp]
 \begin{center}
   \FigureFile(80mm,80mm){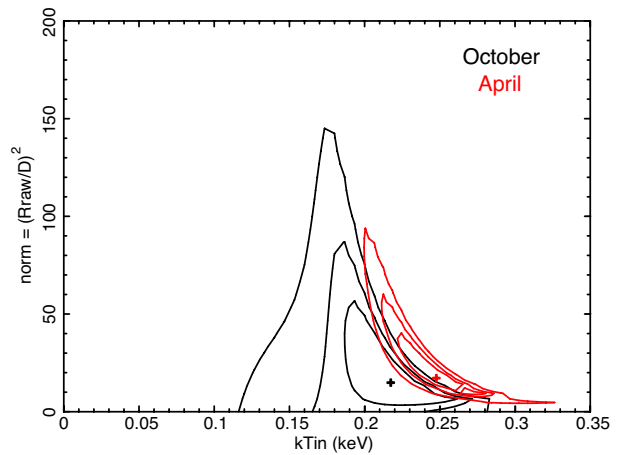}
   \end{center}
    \caption{Confidence-level contours of the ID 2 fits on the $T_{\rm{in}}$ vs $(R_{\rm{raw}}/D)^{2}$ plane. The April data are in red, while the October in black. Crosses indicate the best-fit values in the individual observations, whereas solid lines represent the confidence levels at $68\%$, $90\%$, and $99\%$, respectively.}
    \label{fig:contour}
\end{figure}

For the above purpose, figure \ref{fig:contour} presents confidence-level contours of the ID 2 fit on the $T_{\rm{in}}$ vs $(R_{\rm{raw}}/D)^{2}$ plane. Here, $D$ is the distance to Hol IX X-1 in units of 10 kpc, and the dimensionless quantity $(R_{\rm{raw}}/D)^{2}$ is equivalent to the normalization of the seen-through MCD component. Thus, the acceptable parameter regions from the two observations are mutually exclusive, typically at $90 \%$ confidence levels. However, it is difficult to tell which of $T_{\rm{in}}$ and $R_{\rm{raw}}$ varied, because of the small variation amplitude, and of intrinsic coupling between these two parameters. All what we can say is that the October spectrum, compared with that in April, is characterized by a steeper (larger) value of $\Gamma$, and a reduced contribution from the seen-through disk; the former explains the flux increase in figure \ref{fig:apavesp} from $\sim 5$ keV down to $\sim 2$ keV, whereas the latter effect partially compensate for the former below $\sim 2$ keV.

\begin{table*}[htp]
\caption{Summary of simultaneous fits to the April and October spectra.$^{\ast}$}
\begin{tabular}{cccccccccccc}
  \hline
  ID$^{\dagger}$  & Date 	& $N_{\rm{H}}$		& $\Gamma$			& $T_{\rm{e}}$ 	    	& $\tau$		& $T_{\rm{in}}$		& $R_{\rm{raw}}$ 			& $R_{\rm{in}}^{\ddagger}$	& $L_{\rm{disk}}$	& $L_{\rm{X}}$	& red. $\chi^{2}$ ($\nu$) \\
  \hline
  1	& Apr.	& $1.1^{+0.4}_{-0.3}$	& $1.65^{+0.02}_{-0.03}$	& $2.7\pm0.2$		& $\sim 14$	& $0.29^{+0.06}_{-0.05}$	& $910^{+1300}_{-720}$		& $2630$			& $6.3$		& $1.15$			& $1.02 \ (230)$ \\
  	& Oct.	& $1.9^{+0.5}_{-0.6}$	& $1.79^{+0.01}_{-0.02}$	& $3.2^{+0.4}_{-0.3}$	& $\sim 11$	& $0.18^{+0.04}_{-0.02}$	& $3290^{+4790}_{-3160}$	& $9470$			& $12.2$		& $1.30$  		& \\
  \hline
  2	& Apr.	& $1.3\pm0.3$		& $1.67\pm0.02$		& $2.9^{+0.2}_{-0.1}$	& $\sim13$	& $0.25^{+0.04}_{-0.03}$	& $1330^{+1720}_{-1050}$	& $3790$			& $7.0$		& $1.16$			& $1.04 \ (232)$ \\
  	& Oct.	& ...					& $1.76^{+0.02}_{-0.01}$	& 	...				& $\sim12$	& $0.22^{+0.06}_{-0.03}$	& $1240^{+2220}_{-1160}$	& $5660$			& $9.7$		& $1.30$			& \\
  \hline
  3	& Apr.	& $1.8\pm0.3$		& $1.70\pm0.02$		& $3.0\pm0.2$		& $\sim13$	& $0.21^{+0.03}_{-0.02}$	& $2610^{+2670}_{-2060}$	& $6370$			& $10.2$		& $1.15$ 		& $1.13 \ (234)$ \\
 	& Oct.	& ...					& $1.76\pm0.02$		& ...					& $\sim12$	& ...					& ... 					& $6930$			& $12.1$		& $1.30$ 		& \\
  \hline
  \end{tabular}
  \begin{flushleft}
  $\ast$: Meanings and units for the individual parameters are all identical with those in table \ref{tb:avefitp}.\\
  $\dagger$: From ID1 to ID3, larger number of parameters are constrained to be the same between the two spectra. See text for details.\\
  $\ddagger$: The inner-disk radius defined in equation \ref{eq:rin}.
  \end{flushleft}
  \label{tb:simfitp}
\end{table*}

\section{Discussion}
\subsection{Modeling of the Spectra}
In the two Suzaku observations, Hol IX X-1 exhibited the PL-like state spectrum which extended up to 11 keV. These spectra were decomposed into a hard ($\Gamma = 1.7\--1.8$) PL continuum with a roll over at $\sim 8$ keV, and a weak soft excess in $\le 1.5$ keV. To quantify the $0.6\--11$ keV spectra, we utilized the MCD+THC modeling, which actually gave acceptable fits to both spectra. Since the peak of the soft excess was at $\sim0.9$ keV, we obtained a relatively low $T_{\rm{in}}$ ($\sim 0.3$ keV) compared to those obtained in the Disk-like state of this source ($T_{\rm{in}}=1.24$; La Parola et al.~2001). Considering further that the PL continuum is a consequence of the THC process, Hol IX X-1 is inferred to harbor a cool ($T_{\rm{e}}\sim 3$ keV) and optically thick ($\tau \sim 12$) corona.

Gladstone et al.~(2009) applied the MCD+THC model to several ULXs to characterize their spectra which are generally concave at $2\--4$ keV and convex at $\ge 5$ keV. Their ULX spectra, all with high-energy roll over corresponding to rather low temperatures ($T_{\rm{e}}=5\--8$ keV), were successfully explained in terms of THC in \textquotedblleft thick and cool\textquotedblright\ coronae. Hol IX X-1 is not an exception, and the present result is generally consistent with those in Gladstone et al.~(2009). 

Some authors (e.g. Vierdayanti et al.~2010, Gladstone et al.~2009) suggested that these cool and thick coronae provide evidence that ULXs are accreting matters at extremely high accretion rates, because similar features are seen in several theoretical studies of super-critical accretion flows (Kawashima et al.~2012, Ohsuga, Mineshige \& Watarai~2007). However, this type of \textquotedblleft thick and cool\textquotedblright\ coronae are being discovered in various accreting objects that are shining considerably below their Eddington limits. For example, Noda et al.~(2011) concluded that the \textquotedblleft soft excess\textquotedblright\ in Active Galactic Nuclei can be interpreted, at least in some cases, as the highest end of the THC continuum; a representative Type I Seyfert, Mrk 509, had a corona with a temperature of $T_{\rm{e}} = 0.49$ keV and an optical depth of 17.6. Another example is a report by Sugizaki et al.~(2013) that the neutron-star binary MAXI J0556-332 harbored a corona of $\tau = 8\-- 10$ and $T_{\rm{e}}=1.5 \-- 3$ keV, during its High/Soft (but still sub-Eddington) state. Thus, these cool and thick coronae cannot be regarded as a unique signature of super-critical accretion flows.

\subsection{Interpretation of the Spectral Variability}
From April to October, the X-ray spectrum of Hol IX X-1 became softer and $14 \%$ more luminous. The spectral fit showed that, in October, the contribution of the seen-through disk component decreased and the photon index $\Gamma$ increased, while the coronal temperature $T_{\rm{e}}$ and the absorption column density $N_{\rm{H}}$ remained unchanged within errors. Since $T_{\rm{e}}$ remained constant, the change in $\Gamma$ must be caused by a slight decrease in the optical depth $\tau$ of the Compton cloud (equation \ref{eq:tau}). 

In order to interpret the observed spectral variability, we follow a picture established by Makishima et al.~(2008), which has been motivated to explain spectral variations of the BHB Cygnus X-1 in the Low/Hard state. Specifically, we consider a clumpy and variable corona covering an accretion disk beneath it. The corona is considered to have some sort of \textquotedblleft holes\textquotedblright\ through which the disk component is directly observable as the soft excess. The rest of the disk emission is Comptonized into the hard continuum. In addition, we assume that the disk is stable, while the opening fraction of the corona varies.

Employing the above scenario, let us interpret the observed spectral changes. Compared to the April data, the October spectrum has a smaller contribution from the seen-through disk. This can be explained as a result of a decrease in the opening fraction of the corona. Specifically, the April data indicates an opening fraction of $1-(1330/3790)^2=0.88$, while that of October $1-(1240/5660)^2 = 0.95$, with a relative increase by $\sim 7 \%$. Thus, the corona became less clumpy and more uniform in October, so that a larger number of seed photons became Comptonized.
Furthermore, the less clumpy corona is expected to have a lower local density, leading to a reduced optical depth. This can explain the steeper PL slope (the larger $\Gamma$) in October.
Thus, the change in clumpiness of the corona can explain the observed spectral variation of Hol IX X-1. Of course, the present object is known to exhibit a wider variety of spectral shapes (Luangtip et al.~2016) than we observed. This suggests that $T_{\rm{e}}$ and the average coronal density are also varying sometimes, independently of the opening fraction.
 
\subsection{Absence of X-ray Reprocessing Features}
Accreting systems frequently exhibit emission and/or absorption lines in their spectra, as well as strong and variable low-energy absorption. These features are considered to arise via X-ray reprocessing process (e.g. photoelectric absorption, fluorescence, and resonance scattering), by surrounding materials including accretion disks, stellar winds from the mass-donating stars, and the donor's photosphere. However, as reviewed in  section 1, the ULX spectra tend to be extremely featureless, and weakly absorbed. For example, as mentioned in section 2, Walton et al.~(2013) derived a stringent upper limit on the emission/absorption lines in the present data sets.

Luangtip et al.~(2016) studied the long term spectral variability of Hol IX X-1 by accumulating data taken with XMM-Newton and Swift. While Hol IX X-1 changed both in its spectral shape and the luminosity ($0.9\--2.1\times10^{40}$ erg s$^{-1}$), its absorbing column density remained constant within errors at $N_{\rm{H}}=1.0\--2.1\times 10^{21}$ cm$^{-2}$ which is consistent with our measurements. Since the Galactic absorption column density toward Hol IX X-1 is $4.1\times10^{20}$ cm$^{-2}$ (Dickey \& Lockman~1990), the intrinsic absorption of Hol IX X-1, including contribution from the host galaxy, is of the order of $\sim 10^{21}$ cm$^{-2}$. This $N_{\rm{H}}$ is much lower and less variable than those in Galactic as well as Magellanic High Mass X-ray Binaries, which vary around $\sim10^{22-23}$ cm$^{-2}$, often depending on the luminosity (e.g. Novara et al.~2011). These properties of low and stable $N_{\rm{H}}$ are considered to apply not only to Hol IX X-1 but also to other ULXs, discriminating them from the known High Mass X-ray Binaries. 

According to some computational studies, massive outflows are launched, due to high radiative pressure, from BHs that are accreting above their Eddington limits (e.g. Kawashima et al.~2012). Such outflows, in turn, are expected to produce strong reprocessing features, including low-energy absorption, fluorescence emission lines, absorption edges (of Fe in particular), and resonant absorption lines. In fact, the Narrow Line Seyfert 1 galaxy 1H 0707-495, which is considered to be accreting at a super-Eddington rate, exhibits deep Fe-K absorption features in the spectrum (Hagino et al.~2016), which are blue shifted with a velocity of $\sim 0.2$ times the light velocity. If ULXs were accreting also at very high rates, similar features that are much stronger than were barely detected (NGC 5408 X-1 and NGC 1313 X-1; Pinto et al. 2016) would be ubiquitously detected in their spectra. Therefore, the featureless nature of ULX spectra may not be easily reconciled with the popular view of super-critical accretion.

One possibility is that such outflows from ULXs are highly ionized under strong X-ray illumination. Then, the features such as photoelectric absorption and fluorescence lines will go away. Furthermore, the outflowing plasma can be optically thick to Compton scattering, to make the spectrum even more featureless. However,  outflows in super-critical objects are expected, in reality, to be $\it{less \ ionized}$ than in sub-Eddington sources, for the following reasons. Generally, the ionization degree of  outflows must be determined mainly by the ratio of the ionizing luminosity $L_{\rm{ion}}$ to the outflow density $n_{\rm{of}}$. As the mass accretion rate $\dot{M}$ gets higher, the $L_{\rm{ion}}/\dot{M}$ ratio would decrease because the accretion flow should become more radiatively inefficient, whereas the $n_{\rm{of}}/\dot{M}$ ratio would increase because a higher fraction of $\dot{M}$ should turn into outflows. These effects must lead to a lower $L_{\rm{ion}}/n_{\rm{of}}$ ratio, and hence to lower ionization.

When such Compton-thick outflows no longer reaches complete ionization, we expect to see Fe-K absorption edges at $\sim 8$ keV, because the cross section for photoelectric absorption by Fe ions in Solar abundance materials is a few times higher, at 7 keV, than the Thomson cross section: the plasma between  us and the Compton photosphere would inevitably produce the Fe-K edge. Indeed, a clear Fe-K edge structure is seen in the RXTE/PCA spectrum of the BHB GRO J1655$-$40 (Shidatsu et al. 2015), acquired when the object was very luminous and presumably launching an ionized Compton-thick wind. Since we do not observe such Fe-K edge features from ULXs, the super-Eddington scenario appears unfavored even considering the photoionization.

\subsection{Possible Mass Estimation}
In the super-critical accretion scenario, it has been frequently argued that the \textquotedblleft soft excesses\textquotedblright\ in the ULX spectra originates from photospheres of the optically-thick outflows and the MCD model is just approximating their spectra (e.g. Middleton et al.~2015). However, in plausible parameter ranges concerned here, the outflows could be optically-thick only to Compton scattering, and not to free-free absorption. Then it is not obvious whether a blackbody-like spectrum can be indeed emitted from such outflows. Furthermore, as we showed in $\S$4.3 and $\S$4.4, the spectrum of Hol IX X-1 is actually explicable with the ordinary disk emission model, and the featureless spectrum argues against the super-critical accretion scenario. 
These facts lead us to estimate the mass of the BH in Hol IX X-1 in a rather \textquotedblleft classical\textquotedblright\ way, relying on the physics of standard accretion disk (Shakura \& Sunyaev~1973). 

Since the inner-radius temperature $T_{\rm{in}}$ and the disk luminosity $L_{\rm{disk}}$ are available from the observation, the Stefan-Boltzmann's law,
\begin{equation}
L_{\rm{disk}} = 4\pi R_{\rm{in}}^2 \sigma T_{\rm{in}}^{4},
\end{equation}
allows us to calculate the inner-disk radius $R_{\rm{in}}$. For a non-rotating BH, the innermost stable circular orbit (ISCO) is equivalent to 3 times the Schwarzschild radius $R_{\rm{S}} = 2GM/c^{2}$. Then, we are able to estimate the BH mass by identifying $R_{\rm{in}}$ with ISCO (Makishima et al.~1986). However our spectral model (Model 3) implies that only a small fraction of $L_{\rm{disk}}$ is visible as the soft MCD component, while the rest is Comptonized into the hard continuum. To estimate $R_{\rm{in}}$, we hence need to add, in quadrature, $R_{\rm{raw}}$ (derived from the MCD component) and that from the region covered by the corona, $R_{\rm{com}}$, as
\begin{equation}
R_{\rm{in}}^{2} = R_{\rm{raw}}^{2} + R_{\rm{com}}^{2} \ ,
\label{eq:rin}
\end{equation} 
(e.g. Kubota~2001, Kubota \& Makishima~2004, Makishima et al.~2008). Here, we assume that the number of photons are conserved in Compton scattering. 

We must note that the strong Comptonization seen in Hol IX X-1 would affect the calculation in equation \ref{eq:rin}, by distorting the spectrum of the accretion disk beneath it. Assuming that the observed inner-radius temperature and luminosity are modified by factors of $\alpha$ and $\beta$ (both $> 0$), respectively, the true inner-disk radius would be obtained by correcting the apparent inner-disk radius $R_{\rm{in}}$ by a factor of $\beta^{2} /\sqrt{\alpha}$. Since the Comptonization increases both the luminosity ($\alpha > 1$) and the temperature ($\beta > 1$) by spectral hardening, it is not obvious whether the correction factor $\beta^{2} /\sqrt{\alpha}$ becomes larger or smaller than unity. In order to evaluate which effect ($\alpha$ or $\beta$) will dominate the other, we need to employ more detailed physical models (e.g., $\tt{diskEQ}$; Kubota \& Done~2016), which is out of the scope of the present paper.

The values $R_{\rm{in}}$ and $L_{\rm{disk}}$ derived assuming $\alpha = \beta = 1$ are shown in table \ref{tb:simfitp}. Since the inclination angle for this system is yet to be known, we assumed a face-on geometry in the calculation. Under conditions of non-rotating BH and $R_{\rm{in}}=$ISCO, the mass of BH in Hol IX X-1 is estimated to be $430 \-- 640 \ M_{\rm{\odot}}$, which is in the intermediate-mass regime. However, in some spectral states of BHBs, this $R_{\rm{in}}=$ISCO assumption does not always hold: good examples are the states called the very high state (VHS) and the low hard state (LHS), which are both characterized by strong Comptonization like the present spectra. According to Tamura et al.~(2012), the representative BHB, GX 339-4, showed series of VHS spectra which can be explained with an accretion disk of which the innermost radius is truncated at $1.3\--2.2$ times ISCO. Yamada et al.~(2008) also found that Cygnus X-1 showed a disk truncation at similar radii ($\sim 2$ times ISCO) in the LHS. Since the typical disk truncation radius in BHBs is thus $\sim3$ times ISCO, we may be allowed to assume that the accretion disk in Hol IX X-1 is also truncated at similar radii. Under this assumption, the BH in this ULX is estimated to have a mass of $140\--210 \ M_{\rm{\odot}}$ and shining at $0.3\--0.4\ L_{\rm{Edd}}$; it still supports the intermediate mass BH scenario. If instead the BH in Hol IX X-1 is rapidly spinning, say, with the dimensionless spin parameter $a^{\ast} = 0.5$, and the disk extends down to ISCO, we obtain a BH mass of $1300\-- 1900 \ M_{\rm{\odot}}$, because in this case the ISCO will be at $R_{\rm{S}}$.

In Kobayashi, Nakazawa, and Makishima~(2016), we performed an independent mass estimation focusing on the transition luminosity between the PL state and the Disk-like state. By analyzing the spectra of several ULXs including Hol IX X-1, the transition luminosity was found to scatter over an order of magnitude ($1.7\times10^{39}\--2.2\times10^{40}$ erg s$^{-1}$) among the ULXs studied. If we can identify the transition luminosity with a particular Eddington ratio just like in the hard-soft transition of ordinary BHBs, the mass of the ULXs should also span a similar range. The estimated transition luminosity of Hol IX X-1 is $\sim 2.2\times 10^{40}$ erg s$^{-1}$. Therefore, Hol IX X-1 is estimated to have at least $\sim 100 \ M_{\rm{\odot}}$ if the ULX in the sample with the lowest transition luminosity (M33 X-8) is assumed to have $\sim10 \ M_{\rm{\odot}}$. Thus, the conclusion of Kobayashi, Nakazawa, and Makishima~(2016) agrees with the present estimation. 

Finally, as touched in section 1, the dramatic detection of the gravitational wave event GW150914 (Abbot et al.~2016) has given an irrefutable demonstration that a BH with $> 60 \ M_{\rm{\odot}}$ does exist, and presumably even abundant in the universe. Such BHs will accrete from companions if they form binaries, and even directly from interstellar medium (Mii \& Totani~2005) as they drift into regions of high densities (Nakamura et al.~2016) such as star-forming sites. All these arguments are considered to strengthen the intermediate-mass BH scenario.  

\subsection{Possible mass-accretion regimes}
Now that we regard ULXs are relatively massive BHs under sub- or trans-critical accretion (not highly super critical), it is of essential importance to examine how their PL-like spectral state studied here corresponds to the know spectral states of BHBs, namely, the LHS, the high/soft state (HSS), the VHS, and the Slim-disk state (SDS), which are realized approximately in the increasing order of the mass accretion rate (Kubota \& Makishima 2004). Among them, the HSS and SDS are readily excluded, because they are characterized by convex thermal disk spectra which are distinct from the Compton-dominated signals of Hol IX X-1. We are thus left with the LHS and the VHS, both dominated by THC continua like in ULXs.

Of the LHS and the VHS, the latter appears much closer to the PL-state of ULXs, for the following reasons. One is that the LHS and VHS are known to emerge at relatively low ($< 0.01 L_{\rm{edd}}$) and high (fraction of $L_{\rm{edd}}$; Kubota \& Makishima~2004) luminosities, respectively, making the latter clearly more appropriate as an explanation of ULXs. The other is the temperature ratio between the corona and the seed-photon source, namely $Q \equiv T_{\rm{e}}/T_{\rm{in}}$ (figure 4 of Zhang et al.~2016; Kobayashi et al.~2016); this quantity is considered to represent a balance between heating by hot ions and cooling by the seed photons, both working on the coronal electrons. in BHBs in the LHS, this parameter becomes large at $Q=10^{2} \-- 10^{3}$ because the corona is very hot ($T_{\rm{e}}>$ several tens keV) and the truncated disk is rather cool at sub-keV temperatures. In contrast, the VHS of BHBs is characterized by $Q=10\--50$ because $T_{\rm{in}}$ is somewhat higher at $\sim 1$ keV than in the LHS, and the corona becomes (presumably under stronger photon cooling) much cooler at a few tens keV (e.g., Kubota \& Done 2004, Hori et al.~2014). Clearly, the condition in ULXs is closer to those in the VHS rather than in the LHS, because we find $Q=3.0/0.23=13$ from Hol IX X-1, and similar values of $Q=10\--20$ from other PL-state ULXs (Kobayashi et al.~2016). Interestingly, neutron-star low-mass X-ray binaries are also found either at $Q<5$ in their HSS or $Q>20$ in their LHS (Zhang et al.~2016), avoiding the region of $Q=5-20$ where ULXs are found in their PL-like state.

In spite of the above similarity between the PL-state of ULXs and the VHS of BHBs, we certainly notice a few points of differences. For example, the value of $T_{\rm{e}} \sim 3$ keV derived from Hol IX X-1 is considerable lower than those in the VHS, typically $\sim 20$ keV (e.g., Kubota \& Done~2004,  ). Furthermore, the spectral slope $\Gamma \sim 1.7$ obtained in the present work is flatter (smaller) than is found in the VHS of BHBs, typically $\Gamma = 2.3$ (e.g., Kubota \& Makishima~2004). The latter difference is attributed to generally higher values of the Compton $y$-parameter, which is $\sim 1.2$ in Hol IX X-1, compared to $\sim 0.5$ in typical BHBs in the VHS. Here. we define the $y$-parameter as 
\begin{equation}
y=4k\left( T_{\rm{e}} - T_{\rm{in}} \right)/m_{\rm{e}} c^2 \times \tau(1+\tau/3)
\end{equation}
where $m_{\rm{e}}$ and $c$ are the mass of an electron and the velocity of light, respectively. Since this $y$ represents the THC strength, we may consider, incorporating our discussion in section 4.3, that the PL-state of ULXs may be reached when the THC in the VHS is more enhanced. It is however yet to be investigated whether such a difference in the THC strength can be naturally attributed to the implicitly assumed difference in the BH mass. It is yet to be clarified how the PL-state and MCD-state of ULXs differ and the transition from the former to the latter has any correspondence in BHBs, e.g., the transition from the VHS to the SDS. If theses attempts do not work, we may have to consider, like Gladstone et al.~(2009), that ULXs are in an accretion state that differs (e.g., in the accretion rate or accretion flow patterns) from those known in BHBs.\\

The authors would like to thank to all the member of the Suzaku Science Working Group. The present research has been financed by JSPS KAKENHI Grant Number 269878.

\bigskip





\end{document}